\documentclass[preprint,12pt]{elsarticle}
\usepackage[utf8]{inputenc}
\usepackage{amssymb}
\usepackage{graphicx} 
\graphicspath{{./Figuras/}}
\usepackage{lineno}
\usepackage[hidelinks]{hyperref}
\usepackage{natbib}

\journal{Journal of non-crystalline solids}

\begin{document}
\begin{frontmatter}


\title{Glass engineering to enhance Si solar cells: a case study of $Pr^{3+}-Yb^{3+}$ codoped tellurite-tungstate as spectral converter}


\author{Maiara Mitiko Taniguchi$^{ad}$}
\author{Vitor Santaella Zanuto$^b$}
\author{Pablo Portes$^b$}
\author{Luis Carlos Malacarne$^b$}
\author{Nelson Guilherme Astrath$^b$}
\author{Jorge Diego Marconi$^c$}
\author{Marcos Paulo Belan\c{c}on$^a$}
\ead{marcosbelancon@utfpr.edu.br}

 \address[$^a$]{Universidade Tecnol\'{o}gica Federal do Paran\'{a} (UTFPR), Câmpus pato Branco, Pós-graduação em processos químicos e bioquímicos, Via do Conhecimento Km 1, 85503-390 Pato Branco - Brazil}
\address[$^b$]{Departamento de F\'isica, Universidade Estadual de Maringá, Maringá, PR, Brazil}
\address[$^b$]{Centro de Engenharia, Modelagem e Ciências Sociais Aplicadas, Universidade Federal do ABC, Santo Andr\'{e}, SP, Brazil}
\address[$^b$]{Departamento de Química, Universidade Estadual de Maringá, Maringá, PR, Brazil}

\begin{abstract}

Spectral converters are known to increase photovoltaic energy conversion by minimizing losses due to fundamental non‐absorption and thermalization processes, and have been suggested to surpass the Shockley-Queisser efficiency limit in single junction solar cells. Here we present a detailed spectroscopic study of photoluminescence in tellurite-tungstate glasses doped and codoped with $Pr^{3+}-Yb^{3+}$ and $Ag$ nanoparticles. The energy transfer mechanisms between $Pr^{3+}$ and $Yb^{3+}$ are discussed based on the near infrared emission under excitation at $442$ nm and on the upconversion emission under excitation at $980$ nm. Fluorescence quenching of $^2 F_{5/2}$ level of $Yb^{3+}$ is observed by increasing the concentration of $Pr^{3+}$, as well as by the addition of $Ag$ nanoparticles. In addition, a discussion on the potential of this glass to increase energy production in spectral converters is presented. The results suggest that the few undesirable energy transfer processes occurring in this material are difficult to be controlled or eliminated properly, resulting in intrinsic losses. This discussion is extended to the potential of glass science to enhance energy production in solar cells, showing that newer designs such as bifacial cells may facilitate the exploration of glasses other than soda-lime for mass production of solar cells. The focus on extending the lifespan by reducing UV induced degradation seems to be a more effective approach than the development of spectral converters for Si solar cells.

\end{abstract}
\end{frontmatter}
\section{Introduction}
\label{sec:intro}
The increasing demand for cheap and affordable solar power has motivated thousands of scientists to develop technologies that could enable solar cells to surpass the Shockley-Queisser (SQ) limit~\cite{Shockley2015}. For example, in single junction silicon solar cells (SC), which share $>90$\% of the world photovoltaic (PV) production~\cite{ITRPV2018,Huang2013a}, the efficiency limit is about 30\%  under typical sunlight conditions. 

At the industry level, on the other hand, monocrystalline silicon PERC (Passivated Emitter and Rear Cell) cells have demonstrated efficiencies as high as $\sim$25\% \cite{Green2016}. This level is indeed a more ``practical limit'' for the efficiency when Auger recombination and the space between cells are considered~\cite{Albrecht2017}. PERC cells will be the largest type of SC at production in 2020 \cite{ITRPV2018,Green2016}, thanks to the capability to be easily integrated in the production process. To overcome this level of efficiency, two basic alternatives have been proposed; the use of multi-junction SC's and the spectral converters for single junction SC's. Three junctions already expand the SQ limit to $\sim$50\%, and an infinite number of junctions would have an efficiency limit of $65\%$~\cite{Albrecht2017}. However, even though multi-junction technology has the highest efficiency already demonstrated, it is often difficult and expensive to build in large scale production~\cite{ITRPV2018}. Perovskite SC's have been developed quite fast in recent years and may enable the mass production of multi-junction SC's in the future~\cite{Albrecht2017,Berry2017}.

Alternatively, spectral converters~\cite{Huang2013a,Day2019,McKenna2017,DelaMora2017a} could potentially be developed to advance Si cells technology in mass production today, just as we have seen with PERC. The idea is to modify the light spectrum that reaches the cell, concentrating light at wavelengths more suitable to be converted in free charges. Considering the solar irradiance (about $1000$~W/m$^2$) reaching the silicon SC, theoretically there are $150$~W/m$^2$ of high energy photons (UV/blue) that could be ``downconverted'' and $160$~W/m$^2$ of low energy photons (wavelength higher than $\sim 1050$ nm) that could be ``upconverted''~\cite{DelaMora2017a}. These energies could be used to increase the power available to single junction silicon cells~\cite{fetlinski2019}.

Rare-earth doped tellurite glasses oftenly exhibit high quantum efficiencies which makes attractive to evaluate their potential as spectral converters. This glasses have been investigated as upconverters by Krishnaiah et al. \cite{Krishnaiah2017}, which showed improvement in SC photocurrent production by converting light of $\sim$1500 nm into $\sim$980 nm with an $Er^{3+}$ doped tellurite-tungstate. Another interesting work by Yang et al. \cite{Yang2014} evaluated an $Er^{3+}$-$Yb^{3+}$ codoped tellurite glass as upconverter placed in the backside of the cell. 

On the other hand, Ag nanoparticles (NP) may enhance the local field of the rare-earth ion site and the energy absorbed by surface plasmon resonance of the electrons in the NP can be transfered to the rare-earth\cite{CHENG2017102,RAJESH2017607,Yusoff2015,Dousti2013,Lakshminarayana2009,Du2015203,Fares2014a,AshurSaidMahraz2013,RezaDousti2013,Singh2010}. 
Lima et al. \cite{Lima2017} and Garcia et al. \cite{Garcia2019} have studied tellurites as downconverters based in $Eu^{3+}$ doping and $Ag$ nanoparticles codoping, where UV/Blue photons are converted to orange photons, which are better suitable to generate current in the silicon SC. 

Recent works have proposed $Pr^{3+}$-$Yb^{3+}$ co-doped materials to convert UV/Blue to near infrared photons ~\cite{xinyang2018,muscelli2018,jianxu2019,Chen2008a,Gao2013a}. By populating the $^3P_n$ excited state of $Pr^{3+}$ ($\sim20000$ cm$^{-1}$), the energy can be transferred to a neighbor $Yb^{3+}$, ideally resulting in two $Yb^{3+}$ at the excited state $^2F_{5/2}$($\sim10000$  cm$^{-1}$)~\cite{xinyang2018,muscelli2018,jianxu2019}. This process can efficiently produce near infrared photons matching the peak of spectral sensitivity of the silicon cell~\cite{Huang2013a}. If absorption, energy transfer, and emission are efficient enough to surpass the losses inside the spectral converter, it could potentially overcome the SQ limit.

This paper presents a spectroscopic investigation of a series of $Pr^{3+} $-doped and $Pr^{3+}-Yb^{3+}$-codoped TeO$_2$-WO$_3$-Nb$_2$O$_5$-Na$_2$CO$_3$ (TWNN) glasses. Our group has investigated this host in the last few years\cite{Belancon2014c,Taniguchi2018a} because among other things it exhibits good thermal and mechanical properties which are desirable in order to produce devices. The effect of $Ag$ nanoparticles in a codoped tellurite-tungstate glass was also investigated, once these particles may improve or deteriorate the downconvertion depending on how they interfere in the local field at the rare-earth site and introduce concentration quenching\cite{Taniguchi2018a}. The optical absorption and photoluminescence spectra were obtained for the UV–Vis–NIR range, and the energy transfer mechanisms between $Pr^{3+}$ and $Yb^{3+}$ are discussed. The results are used to evaluate the potential of TWNN glasses as spectral converters.

\section{Experimental procedure}

TWNN glass samples were prepared as described in details in Ref.~\cite{Taniguchi2018a}, following the relation 72.5TeO$_2$-23WO$_3$-3Na$_2$CO$_3$-1.5Nb$_2$O$_5$ in $mol\%$. The glass was doped with 0.5~Yb$_2$O$_3$ and $x$~Pr$_6$O$_{11}$, where $x$ = 0.01, 0.1 and 0.5~$mol\%$. In addition, a sample was prepared with $x=0.1$~$mol\%$ and 2.0~$mol\%$ of AgN$O_3$, which results in Ag nanoparticles (NP) formation in the glass, as previously reported~\cite{Taniguchi2018a}. All samples were annealed by 30 minutes near the glass transition temperature, just after the quenching, to reduce the mechanical stress.

Raman spectra were obtained with a dispersive microscope (Bruker Senterra). The samples were excited by a laser at 532 nm with 20 mW and the Raman intensity recorded in the spectral range of 50-1542 cm$^{-1}$. The laser beam was focused by a $\times$20 objective lens, and the intensity at the detector was integrated for 2~seconds. The optical absorption coefficient was measured using an UV-Vis-NIR double beam spectrophotometer (Perkin-Elmer Lambda 900). Photoluminescence (PL) spectra in the range 450-1150 nm were obtained with a spectrometer (Thorlabs CCS100 ) under 442 nm HeCd laser excitation, and a spectrofluorimeter (Jobin–Yvon Fluorolog-3 ) with a xenon lamp (450 W) or a nanoLED as the pump source was used to measure the photoluminescence in the range 1200-1700 nm and the upconversion emission in the visible range.

\section{Results}
\subsection{Raman spectra}

\begin{figure}[!htb]
\centering
\includegraphics[scale=0.9]{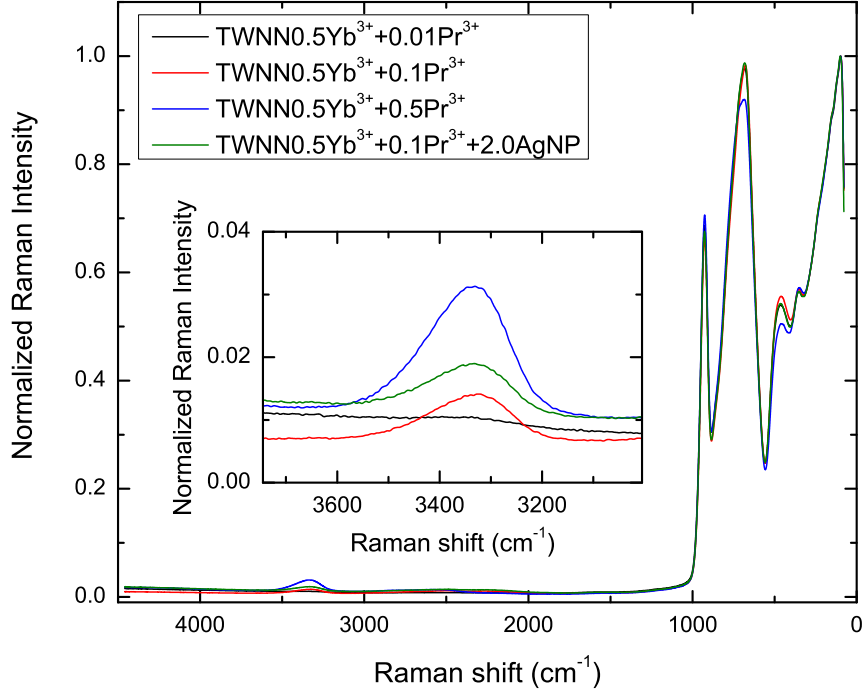}%
\caption{Normalized Raman spectra of TWNN samples. Inset shows the characteristic luminescent peak of Pr$^{3+}$.}
\label{fig:raman}
\end{figure}

The Raman spectra under 532 nm pumping is shown in Fig.~\ref{fig:raman}. The measurements confirmed the expected low phonon energy of the TWNN host, despite of some small differences between the spectra of each sample for energies lower than 1000 cm$^{-1}$. The band at $\sim3350$ cm$^{-1}$ depends on the $Pr^{3+}$ concentration, and the evaluated energy corresponds to the gap between 532 $nm$ and $\sim645$ nm. As we are going to show below, an intense emission is located at this wavelength, suggesting that upper $Pr^{3+}$ levels absorbs at 532 nm, even though this wavelength does not match any $Pr^{3+}$ absorption band. The Pr$^{3+}$ related Raman band increases with dopant concentration and comparing the red and green curves in Fig. \ref{fig:raman}, one can see that $Ag$ codoping has no significant effect.
\subsection{Absorption spectra}

\begin{figure}[!htb]
\centering
\includegraphics[scale=0.9]{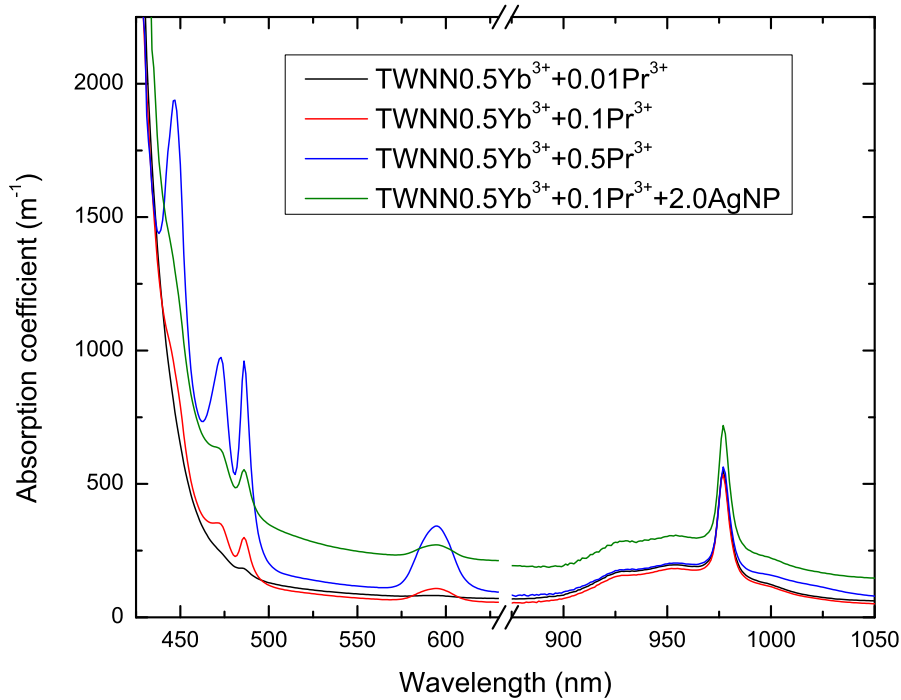}%
\caption{Absorption spectra of TWNN samples in the UV-Vis-NIR range.}
\label{fig:abs}
\end{figure}

Figure~\ref{fig:abs} shows the absorption spectra for all samples. By increasing the $Pr^{3+}$ concentration, the absorption bands between 440-490 nm ($^3P_{0,1,2}$) and around 590 nm ($^1D_2$) become evident. The band at 977 nm is due to the $^2F_{7/2} \rightarrow ^2F_{5/2}$ transition of $Yb^{3+}$. By comparing the absorption spectrum of the sample with $0.1~mol\%$ of $Pr_6$O$_{11}$ with that of the sample with the addition of Ag NP, the absorption coefficient increases in the entire wavelength range, as observed in a previous work~\cite{Taniguchi2018a}.

\subsection{NIR and visible emission spectra}

\begin{figure}[!htb]
\centering
\includegraphics[scale=0.9]{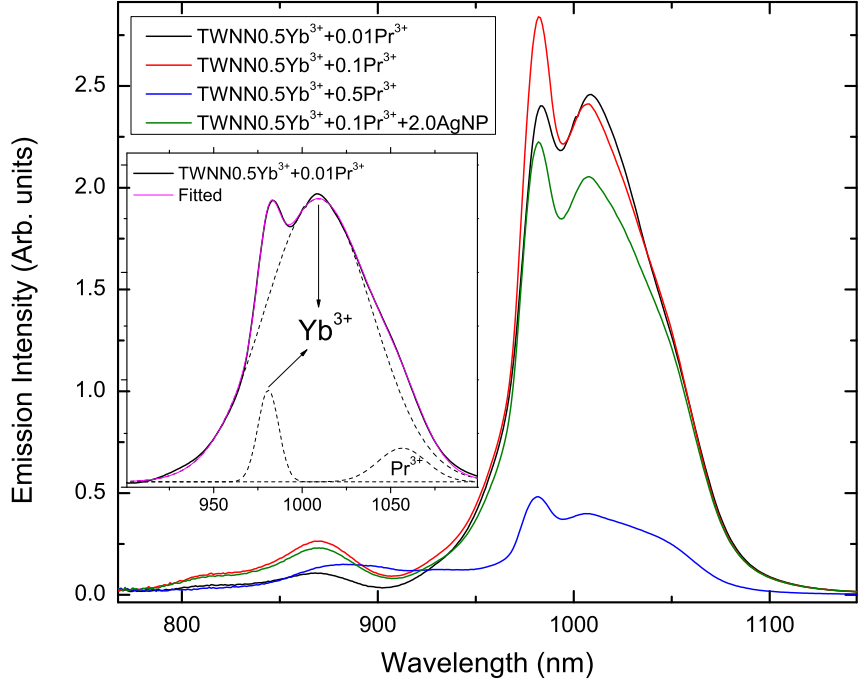}%
\caption{Emission spectra of TWNN samples in NIR range. An intense Yb$^{3+}$ emission is observed around 1000 nm, with the contribution of Pr$^{3+}$ emission at 1050 nm. Inset presents the deconvolution of three radiative transitions.}
\label{fig:YbLumines}
\end{figure}

The near infrared emission spectra under excitation at 442 nm is presented in Fig.~\ref{fig:YbLumines}. Intense emission bands were detected in the range between 950-1050 nm. The radiative transition $^1D_2\rightarrow$ $^3F_{3,4}$ of $Pr^{3+}$ at 1050 nm can be identified by simplified deconvolution of the spectrum of the sample with lower concentration of Pr$^{3+}$ ~\cite{Belancon2014c}, as shown in the inset. A similar result was observed by Muscelli et al. \cite{muscelli2018} in SiO$_2$-Nb$_2$O$_5$ host. The emission bands at 980 nm and 1015 nm are attributed to $Yb^{3+}$ transitions ($^2F_{5/2}\rightarrow$ $^2F_{7/2}$).
As observed previously, the energy is efficiently ($\sim$80\%) transferred from $^1D_2$ level of $Pr^{3+}$ to $^2F_{5/2}$ level of $Yb^{3+}$ in TWNN glass doped with $0.1\%$ of $Pr^{3+}$~\cite{Belancon2014c}. By increasing the $Pr^{3+}$ concentration to $0.5\%$ the $Yb^{3+}$ emission is quenched. The addition of Ag NP also induced a decrease in the emission intensity of $Yb^{3+}$.

\begin{figure}[!htb]
\centering
\includegraphics[scale=0.9]{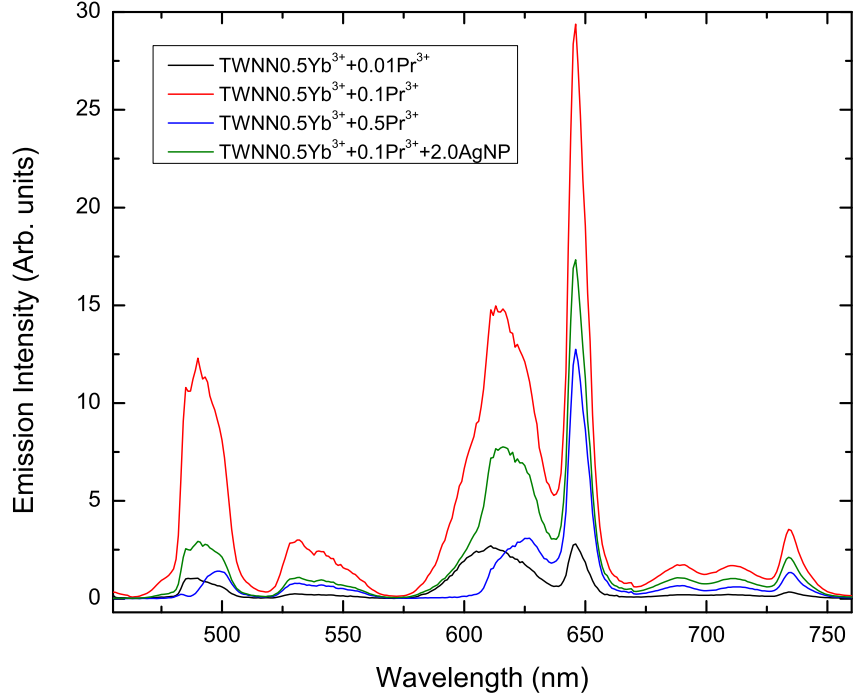}%
\caption{Pr$^{3+}$ emission in the visible range, excited at $442~nm$.}
\label{fig:Prvisible}
\end{figure}

The visible emission spectra are shown in Fig.~\ref{fig:Prvisible}. Emission in the visible region are detected for all samples~\cite{Taniguchi2018a,Belancon2014c}, and can be interpreted as losses in this codoped material, since the main objective is to produce photons matching the peak sensitivity of Si cells. $Pr^{3+}$ self-quenching is effective for Pr$_6$O$_{11}$ concentrations higher than $0.1\%$ in this glass~\cite{Belancon2014c}, reducing primarily the intensity of the resonant transition $^1D_2 \rightarrow ^3H_4$ ($\sim$600 nm). In fact, for the sample with $0.5\%$ of Pr$_6$O$_{11}$ this transition is not present, remaining all the other bands for the decays from the $^3P_{0,1,2}$ excited states, such as $^3P_0~\rightarrow$~$^3H_6$ ($\sim$625 nm) and $^3P_{0,1}~\rightarrow$~$^3F_{2,3,4}$ ($\sim$645~nm)~\cite{Taniguchi2018a}.

\begin{figure}[!htb]
\centering
\includegraphics[scale=0.9]{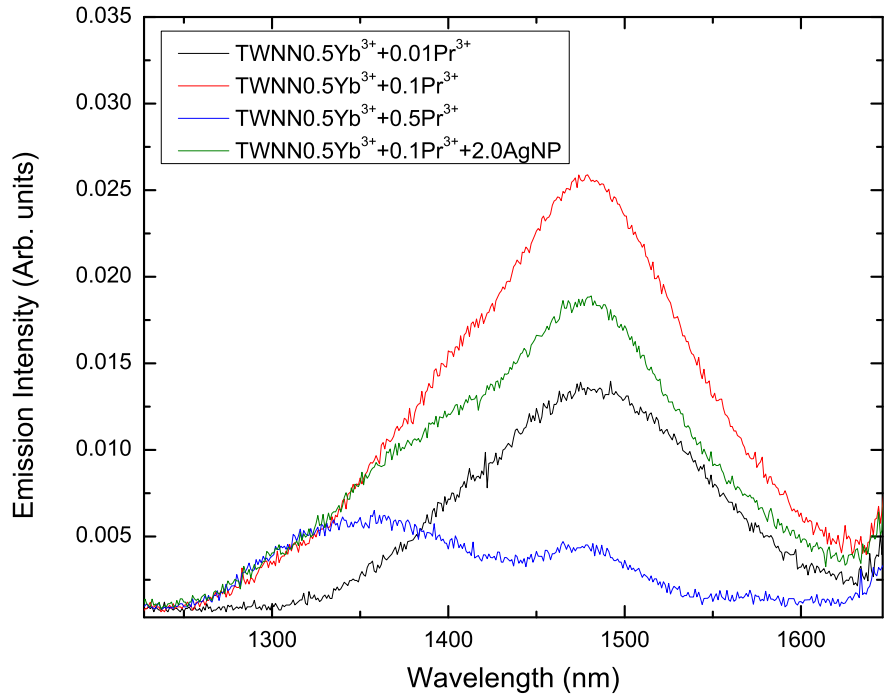}%
\caption{Pr$^{3+}$ emission spectra in the NIR range under excitation at 442 nm.}
\label{fig:cband}
\end{figure}

The effect of Ag NP absorption on the NIR emission spectrum in presented in Fig.~\ref{fig:cband}. The presence of Ag NP quenches the $^1D_2$ level lifetime, and the emission intensity decreases for the $^1D_2~\rightarrow$~$^1G_4$ transition band at $\sim$1480 nm. Recently~\cite{Taniguchi2018a}, we have shown that the effect of Ag nanoparticles in the lifetimes of the $^1D_2$ and $^3P_0$ levels in doped TWNN glasses is difficult to be controlled, possibly because the nanoparticles are being formed or growing below the glass transition temperature in this glass.

\subsection{Upconversion and impurities}
\begin{figure}[!htb]
\centering
\includegraphics[scale=0.9]{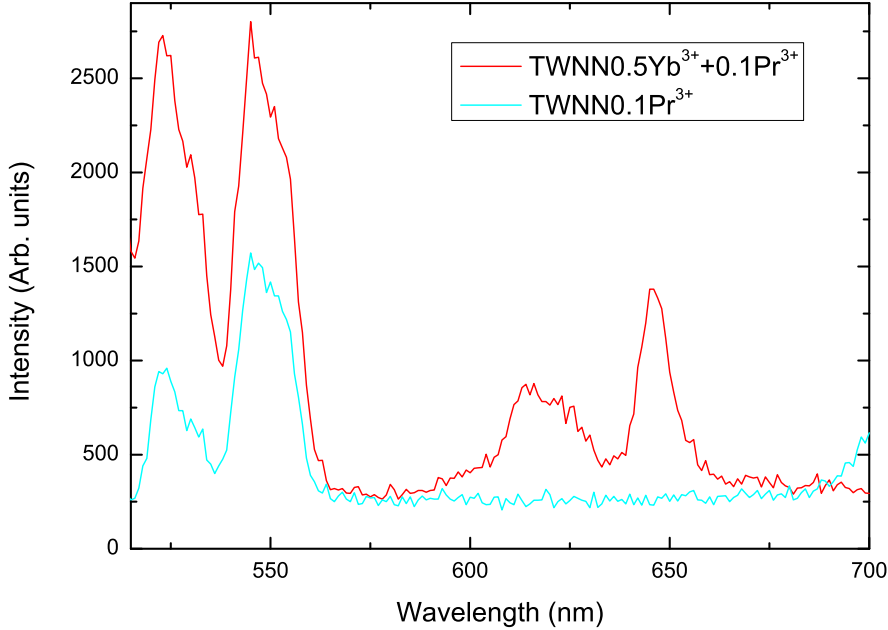}%
\caption{Upconversion emission bands from $980~nm$ photons to the visible range.}
\label{fig:upconversion}
\end{figure}

Figure~\ref{fig:upconversion} shows the upconversion spectra of a $Pr^{3+}$-doped and a $Pr^{3+}$-$Yb^{3+}$-codoped TWNN glasses under excitation at 980 nm. For the $Pr^{3+}$-doped sample, the ground state absorption to the $^1G_4$ level at 1012 nm results in the visible emission around 520-560 nm, which is probably due to transitions $^2H_{11/2}~\rightarrow$~$^4I_{11/2}$ and $^2H_{11/2}~\rightarrow$~$^4S_{3/2}$ of $Er^{3+}$ impurities~\cite{Belancon2014c,Narro-Garcia2013}. Even though the concentration of this impurity is very low, $Er^{3+}$ is well-known by its intense absorption around 980 nm, which is the excitation wavelength in this measurement. When codoped with $Yb^{3+}$, the strong absorption of this ion at 977 nm enhances the upconversion emission from $Er^{3+}$ impurities, and $Pr^{3+}$ transitions between 600-660 nm are observed. This confirms that the energy transfer from $Yb^{3+}$ to impurities and to $Pr^{3+}$ is taking place. Increasing the concentration of any of these ions would indeed increase these losses. As upconversion processes vary non-linearly with the pump power intensity~\cite{Fischer2018}, the effect of such losses should be carefully analyzed in the case of the spectral converter being designed for a concentrated photovoltaic system, in which the power densities used can be hundreds of times higher\cite{Xing2015}.

\section{Discussion}

Since it is easier to convert high energy photons into low energy photons than the opposite, we focuses this discussion on the downconverting potential of TWNN glasses for applications in SC's~\cite{McKenna2017}. Tellurite glasses exhibit very interesting spectroscopic properties~\cite{Jiang2016}. For instance, TWNN glass containing 0.5\%Yb$_2$O$_3$ and 0.1\%Pr$_6$O$_{11}$ can convert blue absorbed photons into infrared emission. However, a path to eliminate most of undesirable Pr$^{3+}$ emissions was not found without affecting negatively the $Yb^{3+}$ NIR emission. The complex energy diagram of $Pr^{3+}$ makes it very difficult to select a transition to be quenched. Although the dopant concentrations might be optimized to enhance the violet-blue to NIR conversion, the effects of other parameters should be carefully evaluated.

In general, tellurites do not have the mechanical strength to withstand real environmental conditions, which makes this class of glasses unlikely to be proposed as substitutes to soda-lime as both cover glass or an external thin layer in SC's. The high refractive index ($\sim2.1$) of the TWNN host~\cite{Taniguchi2018a} results in high reflectivity losses~\cite{Buskens2016}, which may be as high as $13\%$ at the TWNN/air interface. Thus, considering this glass as a coating would require a very specific AR-coating in order to reduce this important loss. Notwithstanding, the availability of tellurium as raw materials is
critical~\cite{Fthenakis2009,Feltrin2007,Davidsson2016,Ali2017,Graedel2011,Grandell2016,Bleiwas2010}.

Commercially, photovoltaic panels are covered by soda-lime glass, which reflects about $4\%$ of the incoming sunlight due to its refractive index of $n\sim1.5$. On the opposite surface of the cover glass, light reaches the encapsulant. As the refractive index of ethylene vinyl acetate copolymer (EVA) matches that of the soda-lime glass, EVA is the most common choice of encapsulant material in Si photovoltaic panels~\cite{Oliveira2018}, reducing the reflection losses while mechanically protecting the cell. Substitutes to EVA, such as Polyolefin, present very similar optical characteristics~\cite{Lopez-Escalante2016} and are already available commercially~\cite{ITRPV2018}.

Replacing soda-lime by TWNN as the front glass would decrease the total irradiance of the cell by $\sim 13\%$ due to reflection losses, i.e., about $\sim130$ W/m$^2$ of the $\sim$1000 W/m$^2$ available. If down-conversion efficiency of $100\%$ could be achieved, it would still mean that half of the $\sim150$ W/m$^2$ available for down-conversion would be lost due to the energy gap between violet/blue and the IR photons produced. This gain would not compensate for the irradiance loss due to the increase of the reflectivity. Some materials exhibiting quantum-cutting with quantum efficiencies well above 100\% were already reported, however only for a narrow range of excitation wavelengths, not absorbing significantly the $\sim$150 W/m$^2$ available. This kind of material would act mostly as a UV protective layer for specific applications~\cite{xinyang2018}.

Special AR-coating could be developed to partially reduce losses intrinsically associated with TWNN glasses. In fact, AR-coatings can increase the total efficiency in commercial Si photovoltaic panels by about 1\% (for example, from 18.0\% to 18.8\%~\cite{Buskens2016}). However, the expected lifetime of such coatings ranges from 10 to 15 years~\cite{ITRPV2018}, decreasing afterwards the amount of light reaching the SC~\cite{Morales2018}. A better alternative would be to eliminate the need of AR-coating by using a glass with refractive index as low as possible. As a consequence, this could simplify the production and recycling process of Si photovoltaic panels~\cite{Xu2018,Lunardi2018}.

The aspects considered in this discussion indicate that the combination of several features could be more effective to increase energy production in SC's than the attempt to surpass the SQ limit, and even though our analysis here was applied to TWNN glass, it is not restricted to this material. Other tellurites have been investigated with particular focus on the description of the quantum efficiencies~\cite{Jiang2016,Han2015,costa2017,HumbertodaCunhaAndrade2018}. However, such specific analysis may lead to incorrect conclusions. For example, Flor\^encio et al. \cite{Kassab2016a} have investigated Zinc-Tellurite glasses covering SC's and, by comparing an undoped to a doped glass, they attributed the better efficiency of the doped glass to the spectral modification produced by the dopant, even though the absorption coefficient spectra for their samples clearly indicate that less light is being transmitted in the undoped sample. Despite the fact that $Yb^{3+}$ emits efficiently in a wide variety of materials, these ions also have a strong absorption between 900-990 nm, and an improved transmission of light in a limited range such as 400-800 nm\cite{Liu2011} does not guarantee an enhanced efficiency.

An interesting perspective is to develop new glasses, possibly by making use of down-conversion~\cite{Green2016a} in order to expand the lifespan of SC's instead of its efficiency. In fact, it is well known that adding some ions such as $Ce^{3+}$~\cite{Oliveira2018} to the front glass results in an enhanced UV resistance and consequently expanded lifespans. This approach could be explored further in future work. For instance, the growing installation of bifacial photovoltaic panels can open the possibility to employ different glass compositions for the back cover of the panels, especially because the strength required for this component is not the same as for the front glass~\cite{Jager2015,Kopecek2018}. At industry level, the share of bifacial Si panels is already expected to grow significantly~\cite{ITRPV2018}, opening the opportunity to develop low cost layers with refractive index near $\sim$1.4, as already demonstrated~\cite{Matsuda2006,Chimalawong2010}. The addition of UV blocking agent, such as $Ce^{3+}$, can help to extend the SC lifespan. However, for this particular ion, the degradation related to iron oxidation inside the soda-lime needs to be bypassed~\cite{Oliveira2018}. Also, potential induced degradation (PID) of SC's is frequently related to $Na^+$ diffusion from soda-lime\cite{Luo2017}, and we believe enhanced glasses or coatings to avoid this mechanism should be evaluated further.

The industry predicts~\cite{ITRPV2018} that Si panels production could reach 1~TW per year, which is almost 10 times the amount produced in 2018 alone. To go beyond such production capacity, new technologies should be developed. Nevertheless, to enhance energy production, it seems reasonable and feasible to extend the average lifespan of Si photovoltaic panels, delaying thus the replacement of panels and increasing the total installed capacity. Considering a scenario where all energy in the world is renewable~\cite{Jacobson2017}, an average of 8~TW electricity from solar would be needed, or about 40~TW of installed capacity. This scale of deployment could be reached by Si photovoltaic panels if the average lifespan is extended to 40 years.

\section{Conclusion}
In conclusion, we have performed UV-Vis-NIR spectroscopic measurements in tellurite-tungstate glasses doped and codoped with $Pr^{3+}-Yb^{3+}$ and $Ag$ nanoparticles. The prospects to continue this research towards the development of a spectral converter were discussed critically, and even though down-conversion has been verified in our samples, the use of TWNN (or any tellurite) as spectral converter seems unlikely to enhance Si photovoltaic efficiency. Even for specific applications, our research indicates that materials with high refractive index do not have a role to play in practice due to the high reflectivity and the low strength often associated with this class of material. On the other hand, our analysis suggest that glass science may still help to develop enhanced SC's, for example, by engineering UV blocking integrated AR-coatings that could increase lifespan by preventing degradation. This would result in higher energy production during the total lifespan of the photovoltaic panels. Such an approach seems feasible, and we are working to contribute to the subject further.

\label{S:4}

\section{Acknowledgments}
\label{S:5}
The authors would like to thank Brazilian agency CNPq (grant $480576/2013-0$) and CAPES for their financial support.




\bibliographystyle{model1-num-names}

\end{document}